\documentstyle[prb,aps,epsf]{revtex} \def\narrowtext{} \tighten \twocolumn
\input epsf.sty

\begin{document}
\draft

\title{Photoelectron Escape Depth and Inelastic Secondaries in High 
Temperature Superconductors}
\author{M. R. Norman$^1$, M. Randeria$^2$, H. Ding$^{1,3}$\cite{HD}, and J. C. 
Campuzano$^{1,3}$}
\address{
(1) Materials Science Division, Argonne National Laboratory, 
Argonne, IL  60439 \\
(2) Tata Institute of Fundamental Research, Mumbai 400005, India \\
(3) Department of Physics, University of Illinois at Chicago, Chicago, 
IL  60607}

\address{%
\begin{minipage}[t]{6.0in}
\begin{abstract}
We calculate the photoelectron escape depth in the high temperature 
superconductor Bi2212 by use of electron energy-loss spectroscopy data.  
We find that the escape depth is only $3 \AA$ for photon energies typically 
used in angle resolved photoemission measurements.  We then use this to 
estimate the number of inelastic secondaries, and find this to be quite 
small near the Fermi energy.  This implies that the large background seen 
near the Fermi energy in photoemission measurements is of some other origin.
\typeout{polish abstract}
\end{abstract}
\end{minipage}}

\maketitle
\narrowtext

Angle resolved photoemission spectroscopy (ARPES) has become one of the 
key tools used to unravel the mystery surrounding the origin of high 
temperature superconductivity in the layered copper oxides\cite{REVIEW}.
The surface sensitivity of this probe is well known, but to date, no
estimate has been made of the photoelectron escape depth.
A related question is the origin of the large 
``background'' present in ARPES spectra near the Fermi energy, which 
needs to be understood before a truly quantitative understanding of the 
data is possible.  In this paper, we make use of electron energy-loss 
spectroscopy (EELS) data to calculate both the photoelectron escape depth 
and the resulting inelastic secondaries.  We find that for photon 
energies typically used in ARPES experiments, the escape depth is low
($3 \AA$) implying that the electrons are coming from the top CuO 
layer.  We then calculate the secondary emission and find that it is very 
small near the Fermi energy, in agreement with earlier estimates by Liu 
{\it et al.}\cite{LIU}.  This implies that the large background is not 
due to secondaries.

To calcuate the escape depth and secondaries, we need to know the 
electron loss spectrum.  Fortunately, in the system most studied by 
ARPES, Bi2212, this has been done some time ago by Nucker {\it et al.} 
\cite{NUCKER}.  They determined the quantity $Im(-\frac{1}{\epsilon})$ 
where $\epsilon$ is the dielectric function.  To aid in our numerical 
calculations, we use a simple analytic form to model these data.  We 
represent the data by a sum of three terms\cite{ZIMAN}
\begin{equation}
Im(-\frac{1}{\epsilon}) = \sum_i c_i \frac{\omega\Gamma_i\omega_i^2}
{(\omega^2-\omega_i^2)^2+\omega^2\Gamma^2}
\end{equation}
with $c_i^{-1}=(6.1,2.1,2.9)$, $\omega_i=(1.1,18.5,32.8)$, and 
$\Gamma_i=(0.7,13.6,17.0)$ (eV units for $\omega_i$ and $\Gamma_i$).  The 
result is plotted in Fig.~1.  The sharp peak at about 1 eV is the plasmon 
associated with the near Fermi energy band.  The broad double peaked 
structure are plasmons associated with the main CuO valence bands, and
is similar to what is seen in Cu metal\cite{HUFNER}.

The inverse differential path length is given by\cite{HUFNER}
\begin{equation}
\lambda^{-1}(E,E') = \frac{1}{\pi a_0 E} \ln(\frac{1+\sqrt{E'/E}} 
{1-\sqrt{E'/E}}) Im(-\frac{1}{\epsilon(E-E')})
\end{equation}
where $a_0$ is the Bohr radius.  The inverse escape depth is then
\begin{equation}
\lambda_{tot}^{-1}(E) = \int_0^E dE' \lambda^{-1}(E,E')
\end{equation}
In Fig.~2, we plot the escape depth.  From the above equations, we expect 
minima near twice the plasmon energy, and this is indeed what we find, 
with a sharp local minimum at about 2 eV, and a much broader global 
minimum at about 50 eV.  We note that for the photon energies typically 
used in ARPES experiments (19-22 eV), the escape depth is quite short 
($3 \AA$).

We now turn to the calculation of inelastic secondaries, due to primary 
photoelectrons which lose energy as they transport out of the crystal.  
The contribution to the photocurrent is\cite{HUFNER}
\begin{equation}
B(E)=\int_E^{\infty} dE' \lambda_{tot}(E) \lambda^{-1}(E',E) I(E')
\end{equation}
where $I(E)$ is the measured photocurrent.  In Fig.~3, we plot this for 
a spectrum at the $(\pi,0)$ point for a $T_c$=87K overdoped Bi2212 sample.  
We compare this to the phenomenological Shirley background often used for 
estimating secondaries which is of the form\cite{LIU}
\begin{equation}
B(E)=c_{Sh}\int_E^{\infty} dE' F(E')
\end{equation}
where $F(E)$ is the primary spectrum ($I(E)-B(E)$).  Fitting to 
the high binding energy tail, we find $c_{Sh}$=0.065, close to the value 
obtained by Liu {\it et al.}\cite{LIU}, and a factor of 25-60 times smaller
than estimates which attribute the flat background of the near Fermi energy 
band to secondaries.  We note the close match of the phenomenological 
Shirley background to the exact result, and thus confirm the earlier 
conclusions of Liu {\it et al.}\cite{LIU}.  That is, the flat background 
cannot be due to secondaries, but is of some other origin.

In conclusion, we have calculated the photoelectron escape depth and the 
resulting secondary emission for the high temperature cuprate 
superconductor Bi2212 using EELS data.  The result is that the escape 
depth is short ($3 \AA$) for typical photon energies, and that the 
resulting secondary emission is too weak to account for the flat 
background associated with the near Fermi energy emission.  This 
background is therefore of some other origin, which we hope to explore in 
a future paper.

This work was supported by the U.S. Dept. of Energy, Basic Energy 
Sciences, under Contract No. W-31-109-ENG-38, the National Science 
Foundation Grant No. DMR 9624048, and Grant No. DMR 91-20000 through the 
Science and Technology Center for Superconductivity.  We also acknowledge 
support from the Aspen Center for Physics, where this work was written up.

\begin{figure}
\epsfxsize=3.0in
\epsfbox{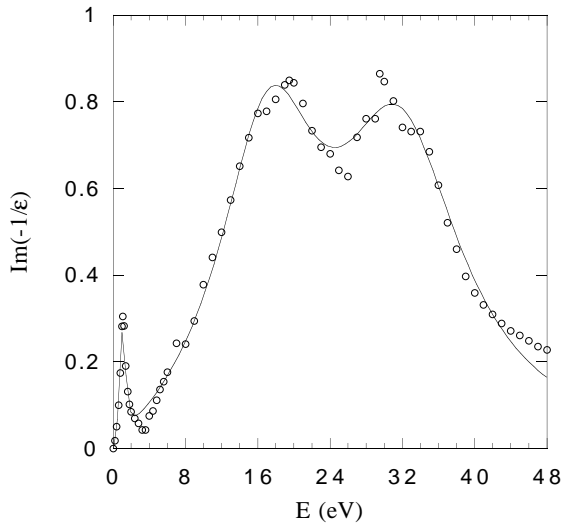}
\vspace{0.5cm}
\label{fig1}
\caption{Plot of the imaginary part of the inverse dielectric function for
Bi2212.  The circles are data of Ref.~3, the solid line the model of Eq.~1.}
\end{figure}

\begin{figure}
\epsfxsize=3.0in
\epsfbox{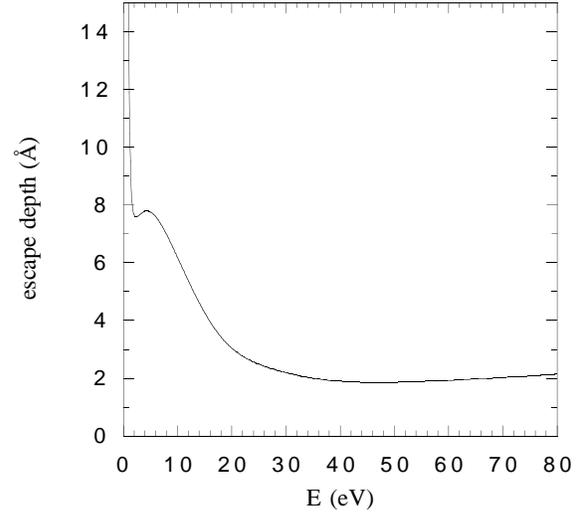}
\vspace{0.5cm}
\label{fig2}
\caption{Plot of the photoelectron escape depth obtained from Fig.~1.}
\end{figure}

\begin{figure}
\epsfxsize=3.0in
\epsfbox{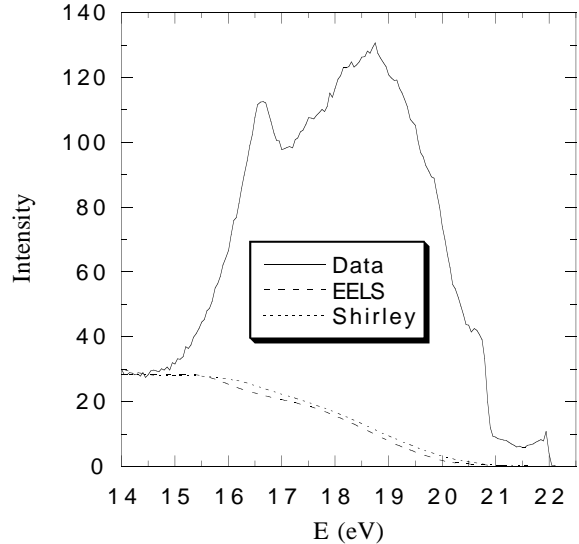}
\vspace{0.5cm}
\label{fig3}
\caption{Main valence band spectrum at the $(\pi,0)$ point for a 
$T_c$=87K overdoped Bi2212 sample (22eV photons).
The dashed line is the calculated 
secondary emission.  The dotted line is the secondaries estimated from a 
phenomenological Shirley background.}
\end{figure}


\begin{references}

\bibitem[*]{HD}Present Address: Dept. of Physics, Boston College

\bibitem{REVIEW}
Z.-X. Shen and D. S. Dessau, Phys. Rep. {\bf 253}, 1 (1995); M. Randeria 
and J. C. Campuzano, in {\it Proceedings of the International School of 
Physics ``Enrico Fermi'', Varenna, 1997} (North Holland, New York, to be 
published).

\bibitem{LIU}
L. Z. Liu, R. O. Anderson, and J. W. Allen, J. Phys. Chem. Solids {\bf 
52}, 1473 (1991).

\bibitem{NUCKER}
N. Nucker {\it et al.}, Phys. Rev. B {\bf 39}, 12379 (1989).

\bibitem{ZIMAN}
J. M. Ziman, {\it Principles of the Theory of Solids} (Cambridge 
University Press, Cambridge, 1972), p. 265.

\bibitem{HUFNER}
S. Hufner, {\it Photoelectron Spectroscopy} (Springer-Verlag, Berlin, 
1996) p. 142.

\end{references}
\end{document}